# CLASSIFICATION OF SPREADSHEET ERRORS

Kamalasen Rajalingham, David R. Chadwick & Brian Knight
University of Greenwich, United Kingdom
Email: k.rajalingham@gre.ac.uk Fax: 020 8331 8665

## ABSTRACT

*This paper describes a framework for a systematic classification of spreadsheet errors. This classification or taxonomy of errors is aimed at facilitating analysis and comprehension of the different types of spreadsheet errors. It is far more comprehensive than any presented or published before. The taxonomy is an outcome of a thorough investigation of the widespread problem of spreadsheet errors and an analysis of specific types of these errors. This paper also contains a clear description of the various elements and categories of the classification. It is also accompanied and supported by appropriate examples.*

## 1. INTRODUCTION

A host of publications over two decades have clearly described the seriousness of spreadsheet errors and their adverse or potential impact on businesses. A financial model review by KPMG Management Consulting, London [1] confirms the frequency and seriousness of spreadsheet errors. Their report states that in 95% of the financial models reviewed, at least 5 errors were found. The review also reveals alarming statistics pertaining to defects in spreadsheet development, addressing the project management, technical and analysis aspects.

An article in *New Scientist* [2] has reported that a study conducted by the British accounting firm Coopers & Lybrand found errors in 90% of the spreadsheets audited. This is an extremely high figure and if the errors went undetected, it could have had a devastating effect on the business. It is evident from these cases that the occurrence of spreadsheet errors is a major problem for businesses and needs to be addressed urgently.

A thorough review of literature relevant to spreadsheet development and errors reveals that very little research has been done on studying specific errors that occur in spreadsheets. The outcome of a thorough analysis of specific types of spreadsheet errors from a wide variety of sources is a more comprehensive classification or taxonomy of spreadsheet errors than ever presented before. It reflects an improvement to the version of the classification presented previously by the authors.

## 2. THE SPREADSHEET ERROR TAXONOMY

### 2.1 Introduction

In a broad sense, taxonomy is the science of classification, though more strictly, it refers to the classification of living and extinct organisms. The term is derived from the Greek *taxis* ("arrangement") and *nomos* ("law"). It is important to note, however, that there is no special theory which lies behind modern taxonomic methods [3].

In attempting to define *taxonomy* within the context of spreadsheet errors, it would be appropriate to investigate the definition of this term in other fields of study. In biology, *taxonomy* refers to the establishment of a hierarchical system of categories on the basis of presumed natural relationships among organisms. The goal of classifying is to place an organism into an already existing group or to create a new group for it, based on its resemblances to and differences from known forms. To this end, a hierarchy of categories is recognised [3].

### 2.2 Taxonomy of Spreadsheet Errors

Based on the definitions borrowed from other disciplines, we can extend the concept of taxonomy to include the classification of spreadsheet errors. For our purposes, the *spreadsheet error taxonomy* can be described as the hierarchical system of categories of spreadsheet errors on the basis of presumed common characteristics and relationships.

Based on the principles of classification adopted in botany and zoology, taxonomic methods for spreadsheet errors depend on:

a) Obtaining a specific type and example of a spreadsheet error

b) Comparing the error with the known range of variation of spreadsheet errors
c) Correctly identifying the error if it has been described, or preparing a description Showing similarities to and differences from known categories, or, if the error is of a new type, assigning it to a new category.
d) Determining the best position for the error in the existing classifications and determining what revision the classification may require as a consequence of the new discovery

## 3. RATIONALE FOR DEVELOPING A SPREADSHEET ERROR TAXONOMY

There are various reasons for developing a taxonomy of spreadsheet errors. The most important probably is that it forces us to clearly understand the characteristics of an error as well as the nature of its occurrence. A comparison can also be made with other related errors belong to the same category or level.

An insight into the features and nature of an error is critical for any effort to devise a solution or method of detection. In general, a similar approach can be taken to address errors within the same category of the classification. The knowledge of the characteristics of an error also enables analysis of its potential impact and frequency. It is also highly probable that other errors in the same category would have the same degree of seriousness.

## 4. FRAMEWORK FOR THE CLASSIFICATION OF SPREADSHEET ERRORS

Two different approaches to the classification of spreadsheet errors were experimented. The following frameworks for the classification of spreadsheet errors appear feasible based on an examination of the process of spreadsheet development [4] and the characteristics of spreadsheet errors and the nature of their occurrence:

i)    Based on the nature and characteristics of the error
ii)   Based on the spreadsheet development life cycle

Having used both frameworks, it was found that the classification based on the characteristics of the error was far more appropriate due to its structure and rigidity. The main criterion for selecting the better framework was the possibility of minimising the recurrence of the same category or type of error in different parts of the taxonomy. In other words, to minimise the overlap of different categories of spreadsheet errors.

In order to produce the taxonomy of spreadsheet errors, the *binary tree* approach is used in conjunction with the analysis of spreadsheet errors based on their nature and characteristics. At each stage of the taxonomy, this approach uses *dichotomies* or divisions into two disjunctive groups, to classify spreadsheet errors.

## 5. CLASSIFICATION OF SPREADSHEET ERRORS

Figure 1 shows the model of the classification of spreadsheet errors constructed by adopting the framework described in the previous section.

SYSTEM-GENERATED

*System-generated errors* are errors made by the spreadsheet software or bugs in the software. Their occurrence is generally beyond the control of users, although they can, when aware, take corrective action.

*Example:*    *Century Error*

In MS Excel 97 for instance, for any entry of a date (without the century) before 01/01/30, the century is assumed to be the 21st century while for any entry of a date (without the century) after 01/01/30, the century is assumed to be the 20th century. This problem of course, can be avoided if the year is explicitly entered with the century e.g. 09/02/1915, 03/12/2060 *(dd/mm/yy)* [5].

USER-GENERATED

*User-generated errors* are errors committed by the user, as opposed to being software/system-generated and can be prevented, detected and corrected by the user. They can be divided into two major categories at the highest level, namely *qualitative errors* and *quantitative errors*.

QUANTITATIVE

*Quantitative errors* are numerical errors that lead to incorrect bottom-line values [6].

**SPREADSHEET ERRORS**

- SYSTEM-GENERATED
- USER-GENERATED
  - QUANTITATIVE
    - ACCIDENTAL
      - DEVELOPER (workings)
        - Omission
        - Alteration
        - Duplication
      - END -USER
        - DATA INPUTTER (Input)
          - Omission
          - Alteration
          - Duplication
        - INTERPRETER (output)
          - Omission
          - Alteration
          - Duplication
    - REASONING
      - DOMAIN KNOWLEDGE
        - REAL-WORLD KNOWLEDGE
        - MATHEMATICAL REPRESENTATION
      - IMPLEMENTATION
        - SYNTAX
        - LOGIC
  - QUALITATIVE
    - SEMANTIC
      - STRUCTURAL
      - TEMPORAL
    - MAINTAINABILITY

**Figure 1. Taxonomy of Spreadsheet Errors**

ACCIDENTAL

*Accidental errors* are mistakes and slips caused by negligence, such as typing errors. Though quite frequently occurring, they have a high chance of being spotted and corrected immediately by the person committing the error. Some, however, do go undetected and could lead to incorrect values in other cells. It is important to state here that most of the errors described under this category can also be intentional or deliberately caused with malicious intent.

After a close examination of various types of accidental errors, it has been found that they can be further divided into two distinct categories. They are *developer-committed errors* and *end-user-committed errors.*

DEVELOPER-COMMIITED ERRORS

*Developer-committed errors* are errors produced by the developer of the spreadsheet model. These errors usually occur in the workings (as opposed to input or output) section of the model. They can belong to any of three categories, namely *omission, alteration* and *deletion.*

OMISSION

Here, omissions are things accidentally left out of the model by the developer. Human factors research has shown that on-fission errors are especially dangerous, because they have low detection rates [6]. It could be that a key factor or variable is omitted from the spreadsheet model and therefore, an important relationship is missing from the model.

*Example:* *References to corresponding input data in the workings/output section are omitted from the model.*

KPMG, in one of their client models, found that increase in vehicle cost was blank until 2001, even though the source of data from that date (from another worksheet) contained values for the earlier years [7].

ALTERATION

This error occurs when the developer of the model accidentally makes a change to an existing model, that produces a defect in the model. An example of such an error is the use of cell protection on the wrong cells accidentally, making it impossible for users to enter data [8].

DUPLICATION

The developer of the model accidentally re-creates elements of the model, causing data duplication or redundancy.

*Example:* *A variable is defined twice.*

When developing a model, it's easy to make a forecast for a growth rate of X%. X is written into the equations that compute growth but is written in as a constant, for example, =[cell above] x 1.04. In a later stage of model development, the user might do a what-if analysis and writes an equation such as = [cell above] x [growth rate cell]. During debugging, the two growth rates might be identical or similar. During use, they might be different [9].

END-USER-COMMITTED ERRORS

*End-user-committed errors* are mistakes or slips made by end-users that merely manipulate or interpret the spreadsheet model/system. The end-users can consist of two distinct groups, namely the *data inputters* and the *data interpreters.*

DATA INPUTTER

The data inputter is the end-user who enters the data required by the model. It is these values which are fed into the *workings* and *output* sections. The data inputter may also produce errors as a result of omission, alteration or duplication of data.

OMISSION

These errors are typically caused by the data inputter who fails to enter a piece of data required by the spreadsheet model.

ALTERATION

These errors usually take the form of data input or overwriting errors. These are errors made by users while adding to or modifying existing data in the spreadsheet model.

*Example:* *Rows are added to spreadsheets but not the "bottom line" totals.*

The modeller has written an equation to find column totals, writing the equation in row seven. Data are to be entered below. The equation is written =SUM (B8:B99). It works fine until a user adds data in row 100. Because this row is beyond the range of the equation, the data is not included in the addition [9].

DUPLICATION

Duplication errors by data inputters are mainly caused by accidentally re-entering data in the wrong part of the spreadsheet.

DATAINTERPRETER

The data interpreter is the end-user who extracts useful information from the model and presents it in a more convenient form. This is the output section of the spreadsheet model. The data interpreter may perform various actions to obtain the desired information. In the process, they may commit errors that can be classed as either omission, alteration or duplication based.

OMISSION

The data interpreter accidentally leaves out certain elements from the output section of the model.

ALTERATION

The data interpreter may incorrectly alter the model and consequently misinterpret the results. For instance, they may sort particular columns of data in a table, accidentally leaving out the corresponding columns. This makes the table inconsistent and unreliable.

ERRORS IN REASONING

These errors involve entering the wrong formula because of a mistake in reasoning. The formulae may be incorrect as a result of either choosing the wrong algorithm or creating the wrong formulae to implement the algorithm.

DOMAIN KNOWLEDGE

*Domain knowledge errors* are produced due to lack of knowledge required to analyse the business function in order to design the data model which is to be represented electronically by the spreadsheet model. These skills enable the user to identify business functions which are suitable for modelling with a spreadsheet and how this modelling is to be done. This requires thorough knowledge of business functionality and requirements for both the present and the future.

REAL-WORLD KNOWLEDGE

These errors involve creating an incorrect formula by selecting the wrong algorithm.

*Example:*      *Calculation of depreciation*
The reducing balance method is used instead of the straight line method or vice versa.

*Example:*      *Absence of distinction between leap and non-leap years*
For instance, year 2000 is a leap year, but calculations divide by 365 not 366 8.

MATHEMATICAL REPRESENTATION

These errors involve incorrect or inaccurate construction of a formula to implement a correctly chosen algorithm.

*Example:*      *The PERCENTAGE problem*

This error occurs when the formula to calculate percentage is incorrectly written, either due to lack of knowledge of what a percentage is or BODMAS (Brackets, Of, Division, Multiplication, Addition, Subtraction) by which the spreadsheet identifies precedence in calculations e.g. **B2/A2*100, B2*100/A2 or B2*A2/100** instead of **A2/B2*100 or A2*100/B2.** This is based on figure 2 below.

| A | B | C | |
|---|---|---|---|
| **Night Wages £** | **Total Wages £** | **Night Wages %** | 1 |
| 1400.00 | 4690.00 | | 2 |

**Figure 2**

IMPLEMENTATION

*Implementation errors* are produced due to lack of knowledge on the full use of the functions and capabilities of the particular spreadsheet package in use, with an understanding of the spreadsheet principles, concepts, constructs, reserved words and syntax. Implementation errors can be divided into *syntax* and *logic* errors.

SYNTAX ERRORS

A syntax error occurs when the formula contains characters and symbols which are not recognised by the spreadsheet software to perform the desired function. Syntax errors can be easily detected as the spreadsheet immediately indicates that an error has occurred.

# LOGIC ERRORS

A logic error is a form of implementation error which occurs when the formula is incorrectly constructed due to a lack of understanding of the features and functions of the spreadsheet software in use. As a result, the formula produces a wrong value.

*Example:     Relative and absolute copy problem*

The relative copy causes cell references in a copied formula to alter row and column references relative to the original cell copied. People often make the false assumption that the software will automatically adapt the cell references wherever they happen to copy '0.

*Example:     Misconception of the AVERAGE function*

Users see the word 'Average' in the column heading and immediately apply the average function without questioning whether it was appropriate 10. Based *on figure 3,* Over 80% of students in a survey entered **=AVERAGE(C6:D6)** in cell F6. But this gives the average of *Basic Wages* and *Overtime Wages* when, given the context, surely it is the 'average wage per person' and the formula should be **=E6/B6.**

|  | A | B | C | D | E | F |
|---|---|---|---|---|---|---|
| 1 | **Lazy Days Staff Budget Costs 1995-1996** | | | | | |
| 2 | | **Staff** | **Basic** | **Overtime** | **Total** | **Average** |
| 3 | | **Numbers** | **Wages £** | **Wages £** | **Wages £** | **Wage £** |
| 4 | | | | | | |
| 5 | **Managers** | **1** | **17700** | **0** | | |
| 6 | **Grade 1** | **3** | **45540** | **1400** | | |
| 7 | **Grade 2** | **9** | **122340** | **2000** | | |
| 8 | **Grade 3** | **12** | **102350** | **0** | | |
| 9 | **Grand Totals** | **25** | **287930** | **3400** | | |
| 10 | | | | | | |

**Figure 3**

*Example:     Circular references*
This error frequently occurs in totals where the formula uses its own value in its calculation. This error will give a run-time error message and so probably occurs infrequently. A common example of a circular reference arises when calculating bank overdraft interest, and can be corrected as follows [8]:

With a circular reference, i.e., the *incorrect* way:

| Cashflow | £ |
|---|---|
| Opening bank balance (overdrawn) | (x) |
| Add: Receipts | x |
| Less: Payments | (x) |
| Less: Overdraft interest based on closing balance | (x) |
| Closing bank balance | (x) |

**Figure 4a**

Each time the spreadsheet is recalculated the overdraft interest will change and update the closing bank balance ad infinitum. Without a circular reference, i.e., the *correct* way:

| Cashflow | £ |
|---|---|
| Opening bank balance (overdrawn) | (x) |
| Add: Receipts | x |
| Less: Payments | (x) |
| Balance before overdraft interest | (x) |
| Less: Overdraft interest on balance before interest | (x) |
| Closing bank balance | (x) |

**Figure 4b**

QUALITATIVE

*Qualitative errors* are errors that do not immediately produce incorrect numeric values but degrade the quality of the model. The model also becomes more prone to misinterpretation on the part of the user. As a result, it also becomes more difficult to update and maintain the model. A more detailed investigation into qualitative errors reveals that they can be generally divided into two different types, namely, *semantic* and *maintainability errors.*

SEMANTIC ERRORS

Semantic errors are qualitative errors that occur due to a distortion of or ambiguity in the meaning of data. It consequently leads to incorrect decisions, choices or assumptions. As far as qualitative errors are concerned, semantic errors are relatively very difficult to detect. They can be divided into *structural* and *temporal* errors.

STRUCTURAL ERRORS

These errors usually take the form of flaws in the design or layout of the model, incorrect or ambiguous headings, and situations in which the documented assumptions are not reflected in the model, causing confusion.

Example:      Formatting error

If you format to one digit to the right of the decimal (F1), and then enter values having greater precision, the spreadsheet will round off the numbers. Thus 1.44 will round off to 1.4; the sum of 1.44 and 1.44 will round to 2.9 from 2.88. Such additions will appear incorrect [9].

*Example:      SUM Incorrect Use Problem*

A common error is to enter any formula within the SUM brackets as though the SUM was mandatory for defining a formula, for instance, in the spreadsheet model in figure x, the formula in cell H7 might be wrongly entered as **=SUM(G7/D7)** when it should really be **=G7/D7**. Although the calculation is correctly done, this is logically wrong and could cause confusion [10].

TEMPORAL ERRORS

Temporal errors are described as qualitative errors produced due to the use of data which has not been updated. They can lead to unreliable decisions or interpretation of the situation.

*Example:      Qualitative error resulting from the referencing of non-current Data*

This is an example of a qualitative error produced as a result of referencing a piece of data that has become invalid due to time lapse. In the example given below *(figure 5),* this piece of data is the exchange rate from British Pounds (£) to Ringgit Malaysia (RM) contained in cell F2. If the exchange rate undergoes acute fluctuations and the changes are not reflected in cell F2, the calculation in cell A8 produces a value that is invalid. This is a qualitative error and any decision made based on this value would be unreliable.

| A | B | C | D | E | F | |
|---|---|---|---|---|---|---|
| | Tea (£) | Milk(£) | Coffee (£) | | Exchange Rate (£ to RM) | 1 |
| 1st Quarter | 450 | 560 | 467 | | 7.3 | 2 |
| 2nd Quarter | 904 | 900 | 352 | | | 3 |
| 3rd Quarter | 872 | 800 | 233 | | | 4 |
| 4th Quarter | 123 | 234 | 901 | | | 5 |
| | | | | | | 6 |
| Total Sale of Tea & Coffee (RM) | | | | | | 7 |
| **=SUM(B2:B5,D2:D5)*F2** | | | | | | 8 |

**Figure 5**

MAINTAINABILITY

Maintainability flaws are features of the spreadsheet model that make it difficult to be updated or modified. They can potentially cause inconsistency in the model. A common and typical example of a maintainability error is hard-coding.

Example:    *Hard-coding*

The hard-coding of a formula is another example of a qualitative, decision error. This error decreases the quality of the spreadsheet by making it much less flexible. Referring to *figure 6,* if the formulae in column H were hard-coded e.g. **=G8/9** (in cell **H8)** instead of **=G8/D8,** and if any of the values in column D (number of staff) changed, the formula in column H of the same row would have to be re-written. This is just a simple example to illustrate the concept of hard-coding being a source of error.

**Staff Budget Costs 1995-1996**

| C | D | E | F | G | H | |
|---|---|---|---|---|---|---|
| | **Number of Staff** | **Day Wages £** | **Night Wages £** | **Total Wages £** | **Average Wage £** | 5 |
| **Grade 1** | 1 | 17700.50 | 0.00 | **=SUM(E6:F6)** | **=G6/D6** | 6 |
| **Grade 2** | 3 | 45540.00 | 1400.55 | **=SUM(E7:F7)** | **=G7/D7** | 7 |
| **Grade 3** | 9 | 122340.00 | 2000.00 | **=SUM(E8:F8)** | **=G8/D8** | 8 |
| **Grade 4** | 12 | 102350.25 | 0.00 | **=SUM(E9:F9)** | **=G9/D9** | 9 |
| **Grand Total** | **=SUM(D6:D9)** | **=SUM(E6:E9)** | **=SUM(F6:F9)** | **=SUM(G6:G9)** | **=G10/D10** | 10 |

**Figure 6**

It should also be noted that some numbers, which at first sight appear to be constants, are often in fact variables. For example, the rate of inflation or the percentage value for employees' pension contributions [8].

**6. CONCLUSION**

The classification of spreadsheet errors has been found to be very useful in analysing specific types of spreadsheet errors. It also enables users to gain a better understanding of the different types of errors that can occur in their spreadsheet models. Appropriate tools, techniques and methods can subsequently be developed to prevent their occurrence in the first place or enhance the chances of detecting these errors after they have occurred. In addition to that, when a new specific type of error is identified, it can be placed in the appropriate category within the taxonomy. In the process of classifying the error, spreadsheet developers and end-users are bound to gain a much deeper understanding of the error. This is because they will be forced to examine and compare its characteristics with those of other spreadsheet errors.

## REFERENCES

(1)   KPMG (1997). *Executive Summary: Financial Model Review Survey,* KPMG
      Management Consulting (London).

(2)   Ward M (1997, August 16). *Fatal Addition,* New Scientist.

(3)   Encyclopaedia Britannica Online

(4)   Rajalingham K, Chadwick D, Knight B and Edwards D (1999). *An Integrated
      Spreadsheet Engineering Methodology (ISEM),* Proceedings of the Third Annual IFIP
      TC- 11 Working Group 11.5 Working Conference on Integrity and Internal Control in
      Information Systems, Amsterdam, The Netherlands, 18-19 November 1999, Kluwer
      Academic Publishers.

(5)   Rajalingham K and Chadwick D (1998). *Integrity Control of Spreadsheet: Organisation
      & Tools,* Proceedings of the Second Annual IFIP TC- 11 Working Group 11.5 Working
      Conference on Integrity and Internal Control in Information Systems, Virginia, USA,
      November 19-20, 1998, pp l47-168, Kluwer Academic Publishers.

(6)   Panko R R and Halverson R P, Jr. (1996). *Spreadsheets on Trial: A Survey of Research
      on Spreadsheet Risks,* Proceedings of the Twenty-Ninth Hawaii International Conference
      on System Sciences, Maui, Hawaii, January 2-5, 1996.

(7)   KPMG (1998).*Review of Client Spreadsheet Models,* KPMG Management Consulting
      (London).

(8)   Batson J and Brown A (1991, autumn). *Spreadsheet Modelling Best Practice,*
      Accountants Digest, 2730.

(9)   *Stang, D. (1987), Spreadsheet Disasters I have known, Information Centre, 3:11,
      November, pp. 2630.*

(10)  Chadwick D, Knight J and Clipsham P (1997). *Information Integrity In End-user
      Systems,* Proceedings of the IFIP TC- 11 Working Group 11.5 First Working Conference
      on Integrity and Internal Control in Information Systems, Zurich, Switzerland, December
      1997.